# Practical Speech Recognition with HTK

Zulkarnaen Hatala
*Jurusan Teknik Elektro*
*Politeknik Negeri Ambon*
Ambon, Indonesia
dzulqarnaenhatala@gmail.com

*Abstract*— The practical aspects of developing an Automatic Speech Recognition System (ASR) with HTK are reviewed. Steps are explained concerning hardware, software, libraries, applications and computer programs used. The common procedure to rapidly apply speech recognition system is summarized. The procedure is illustrated, to implement a speech based electrical switch in home automation for the Indonesian language. The main key of the procedure is to match the environment for training and testing using the training data recorded from the testing program, HVite. Often the silence detector of HTK is wrongly triggered by noises because the microphone is too sensitive. This problem is mitigated by simply scaling down the volume. In this sub-word phone-based speech recognition, noise is included in the training database and labelled particularly. Illustration of the procedure is applied to a home automation application. Electrical switches are controlled by Indonesian speech recognizer. The results show 100% command completion rate.

## I. INTRODUCTION

Automatic voice recognition system or automatic speech recognition system (ASR) aims to translate the digital signals of the human voice into text or written forms that are meaningful according to certain grammar. The human voice signal is captured by the microphone and processed by the computer to get a regular text. Thus ASR can be used for further purposes such as giving a command to a computer to do a group of tasks by just talking. For example, someone at his home instructs the computer to turn off and turn on the bathroom lights, garden lights without having to press the electrical switch enough to give a voice command. Another case of applying ASR is when someone is driving a vehicle and wants to instruct his smartphone to notify the current position, or destination route information or even information about the on-off condition of all the electrical switches at his home.

Based on the number of words supported, ASR is categorized into small size vocabulary ASR (SVASR) and large vocabulary ASR (LVASR). SVASR only supports very few words compared to LVASR. An example of SVASR is turning on-off electrical switches at home. Another example would be digit recognition applications or instruction of a phonebook contact to dial automatic calls. In contrast, LVASR is used as a keyboard replacement for typing by naturally speaking naturally on search engines for topics and unlimited wording. Even LVASR can be used as an automatic translator machine between two different languages such as between English to Indonesian and vice versa. Here the speaker with an LVASR machine(smartphone) does not need to understand the language of the interlocutor because it will be translated by the device [1].

But LVASR requires relatively high-quality hardware with a powerful processor and large memory. This makes LVASR difficult to apply directly to limited hardware environments such as embedded devices, smartphones and home appliance peripherals. This is where the SVASR has the opportunity to be developed. With a small number of words certainly does not require high hardware prerequisites. SVASR can also be developed quickly with high accuracy. Some of these things make the SVASR get a significant place to translate simple voice commands that are useful for everyday life.

## II. SPEECH RECOGNITION

### A. Noisy ASR

Many reports on ASR research are based on speech corpus that recorded on the sound proof noise-free environment that results in crystal clear speech databases[2][3]. The acoustic modelled achieved from these speech databases significantly suffer low performance when applied directly to the noisy environment[4]. Unfortunately, most applications are deployed in environments that are contaminated by noise and disturbance.

### B. Hidden Markov Toolkit

Hidden Markov Toolkit (HTK) [1] is a group of programs and libraries used to develop ASR with the Hidden Markov model. Actually, HTK itself is complete to build the whole ASR starting from speech database, feature extraction, construction and training of acoustic models and conducting test offline and real-time online.

## III. RESEARCH METHOD

A procedure is proposed to build a small vocabulary automatic speech recognition system (SVASR). In general, these steps are to determine the grammar to be supported, create a database training database, feature extraction, create and train sub-word acoustic, i.e. model training, testing and deployment.

### A. Grammar

In designing the SVASR the first thing is to determine the grammar received by the ASR. Grammar for SVASR differs from grammar from LVASR in terms of the number of words for SVASR is less. Grammar SVASR is simpler with fewer production rules. After the grammar has been determined, the following steps are followed:

*a) Determining word lists:* Based on grammar, all word lists can be found and sorted.

*b) Make a spelling dictionary (subword pronunciation dictionary):* Every word that is listed on the word list is mapped to its subword (phoneme). For a list of phonemes to be used are adapted from the list of the International Phonetic Association (IPA) [5], or TI-MIT [2][6].

B. *Creating voice database*

To create a voice database, the procedures that are carried out are:

*a) Voice recording:* Sound recorded using a microphone that matches the desired conditions and matches the test conditions. In the sense that it avoids differences between training conditions and test conditions. For example, when recording was performed in a quiet laboratory condition, but then testing is carried out in high disturbance conditions, the accuracy of testing will certainly decrease. Also recorded voice carefully designed to refers the word list in step A (grammar) before. By not recording words that are outside the word list, the recording process is certainly faster compared to recording additional unused words. The sampling frequency used is 16KHz [6], and the results are saved as a 16 bit PCM file. In recording, microphone sensitivity is also tuned to produce a good sound database. If the microphone is set too sensitive, the sound that occurs can contain too much noise, which can reduce ASR performance.

*b) Labeling of the voice database:* each recorded file is then marked by subword or phonemes. Softwares used to label voice are Speech Filing System, SFS [7] and Praat [8]. SFS has advantages in terms of compatibility with HTK. Sound files are displayed in the spectral domain, then each phoneme is marked in sequence based on time. Naming phonemes using the International Phonetic Association (IPA) guidance [5] or ARPABET even TIMITBET. While examples of labeling each particular phoneme against its spectral model follow the example of CMU Arctic Speech Databases [9] or TIMIT Database [2]. Recording is also performed upon noise patterns in addition to the words in grammar.

C. *Feature Extraction*

Recorded sound files are then converted to feature format. The format used is Mel Frequency Cepstral Coefficient (MFCC) [10]. After that the 39 coefficients could be filtered using the PCA method [11]. This will increase the recognition accuracy in case the input sound has been distorted by noise [12]. Besides that PCA also has the advantage of reducing computer computation which gains performance on low computing systems and embedded devices.

D. *Acoustic Modeling*

After obtaining feature files in the MFCC format, an acoustic model was constructed in the Hidden Markov Model format with an emission distribution in the form of a Gaussian Mixture Model (HMM-GMM) [13]. This phase is often also called the training phase. The proposed acoustic model is sub-word based. That is one HMM model for a particular phone (sub-word), for example one HMM for /**s**/, one HMM for /**ah**/ and so on.

E. *Testing*

After the acoustic model is obtained, it is time to test the model. In this phase, the level of accuracy or success of an SVASR is measured.

F. *Refinement*

Sometimes after in the field testing online or in real time there are errors. These errors can be corrected if the test data has been recorded. Data from this field can be used to correct the errors. In this case, the system is retrained by including the newly acquired error data.

G. *Deployment*

After the desired results of some performance criterions are achieved, the ASR is planted into the target environment. The deployment environment can be different but possibly the same as the offline and online testing environment. If the testing environment is different, certain adaptation techniques must be performed. Adaptation processes are not needed if the planting environment is exactly the same between the testing environment or the same as the sound database recording environment.

Finally, all the steps above are summarized in Figure-1.

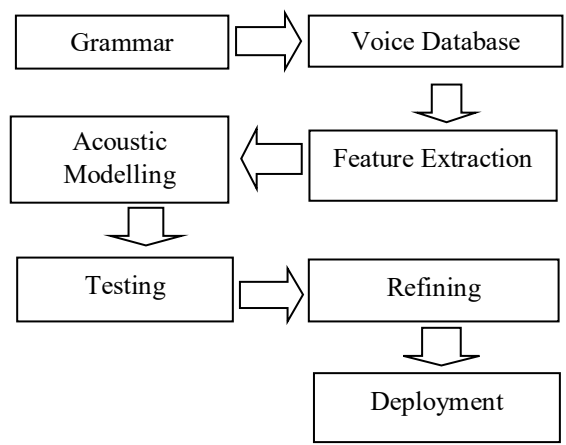

Fig. 1. *Research method for SVASR*

IV. ILLUSTRATION

The following example illustrates the development of SVASR with the proposed procedure. The application that is exemplified is building a recognition system for the Indonesian voice based electrical switcher *(IDSwitch)* and planting it into home automation electrical circuit switches.

A. *Prerequisites softwares*

Besides preparing computer and Bluetooth microphones, supporting software, libraries and applications must be installed as seen in TABLE I. HTK is downloaded and then compiled using Cygwin-gcc or Visual Studio. A slight modification of HTK source prior compilation are required to support raw databases recording, volume scaling and Principle Component Analysis, PCA[11].

TABLE I. SOFTWARES AND DEVELOPMENT LIBRARIES

| No | Software, Libraries | | |
|---|---|---|---|
| | *Software* | *Version* | *License* |
| 1 | Speech Filing System | Release 4.9, SFSWin 1.9 | UCL |
| 2 | HTK | 3.4.5 | Cambridge University |
| 3 | Visual Studio | 2008 | Microsoft |
| 4 | Praat | 6.0.43 | Paul Boersma |
| 5 | Windows OS | Pro 10.0.1439 | Microsoft |
| 6 | Cygwin-gcc | 2.11.2(0.329/5/3) - i686 | GPL |

### B. Grammar

Grammar for IDSwitch is as shown in Fig.2. In this figure the first 4 natural numbers in Indonesian language represent the number of electrical switches for home installation. The *bar '|'* means logical 'OR'.

```
SWITCH_DGT = NUM_0 | NUM_1 | NUM_2 | NUM_3 | NUM_4 ;
SILENCE = SIL ;
SWITCH_W = BATHROOM | GARDEN | AIR_COND;
ON_OFF = ON| OFF|HIDUP|MATI;
CONT = < SWITCH_DGT > ON_OFF | SWITCH_W  ON_OFF;
SIL_NOISE = <SIL | NOISE> ;
SENTENCE = { <$CONT SIL_NOISE> } ;
(SILENCE SENTENCE SILENCE)
```

Fig. 2. *Grammar* for IDSwitch

And for the dictionary of the spelling of the sub-word pronunciation dictionary following the phoneme list in [6] is quoted in part as in Fig.3:

```
NOISE    t sp
NOISE    d sp
NUM_0    k oh s oh ng
NUM_1    s ah t uh
NUM_2    d uw ah
NUM_3    t ih g ah
NUM_4    ah m p ah t
```

Fig. 3. Pronounciation dictionary untuk IDSwitch

So on the dictionary at Fig.3 above the word is placed in the first column and the pronunciation is placed in the next columns. Sub-word **sp** at the end of some entries in the dictionary above stands for *short pause* which is if silence occurs but in a short time. The word NOISE represents noise that might occur during testing. For example, in Fig.3 noise that appears is defined as a pattern of certain phoneme sequences such as the sequence of two consonants without vowels between.

In Indonesian some words appear sequentially as contraction which is the suffix of the word is the prefix of the next word. An example is if the sentence NUM_3 NUM_4 is pronounced then /ah/ is the phoneme suffix of the word NUM_3 at the same time become the prefix to the next word NUM_4. In this case **HVite** often only recognizes one word, which means that word deletion error occurs. The solution used is to create new words in the dictionary. An example is for the case of NUM_3 NUM_4 the solution is like in Fig. 4

```
NUM_3_4    t ih g ah m p ah t
```

Fig. 4. *Contraction modeling in dictionary*

*Contraction* modelling with *dictionary entry* is effective enough for SVASR, but could be a problem when applied to LVASR because of huge entries in order to accommodate all possibilities of contractions.

### C. Speech database

*a) Voice recording:* sound is recorded using HSLab and HVite software with bluetooth microphone input devices. Other recording softwares such as Audacity and CoolEdit is possible, but these softwares are not used, because the desired database is really the one that equal with the test conditions. For bluetooth microphones used are common branded ones and are easily found on the market. This device can be obtained at a low price but without the noise cancellation function. This microphone also contains only one single sensor as according to various literature is less good than those containing an array of censors. An example of a microphone used can be seen in Fig. 5 The advantage of using a wireless microphone is that the speaker is free to move and not tied to an ASR processor, computer or smartphone.

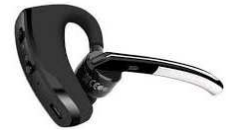

Fig. 5. *Bluetooth Microphone*

The recording place are ordinary housing, office space and laboratory condition with a moderate level of disturbance. Setting microphone's sensitivity also affects the accuracy. For the Bluetooth microphone, the volume is scaled by *c*. Suppose $x(t)$ is the audio signal produced by the microphone then $x^\wedge(t)$ is a signal that is processed by HTK after being scaled according to the equation in Fig. 6.

$$x^\wedge(t)=cx(t), \ 0<c<1$$

Fig. 6. *scaling microphone volume*

This aims to reduce environmental noises encounter such as human steps, animal sounds and possibly the sound of a vehicles engine. Many times, these noises trigger the silence detector of **HVite**, showing that these noises are detected as words from a word list. This reduces the accuracy of SVASR.

As an illustration of the noise parameters, the condition of the room with a common standing fan within 30 cm from the microphone will show an *N* indicator of 35 dB without voice and total (*S+N*) 70 dB with commands voiced. The average power measurements of various recording environments are listed in table II.

The sampling frequency used is 16000 Hz or the period between samples is 625 ns. The recorded file is stored in the HTK format raw 16 integer signed bit.

TABLE II. VARIOUS NOISE CONDITION

| No | Noise parameter under variety conditions | | |
|---|---|---|---|
| | Condition | Measurer | S+N, N (dB) |
| 1 | Living room no fan | HVite | 60, 20 |
| 2 | Living room 30 cm fan from mic. | HVite | 70, 35 |
| 5 | Living room midnight silence | HVite | 70, 15 |
| 5 | Raining (0.3 scaled) | HVite | 43, 15 |

*b) Transcribing (annotating):* Sound is transcribed using Praat [8], then converted into SFS [7] format and converted again into HTK format. This is because Praat cannot convert directly to the HTK format. To label training data, the subword (phoneme) sequences follow the grammar generated in section B. In Figure 7 a sample example of subword with Praat is shown for the order of words NUM_2 NUM_1.

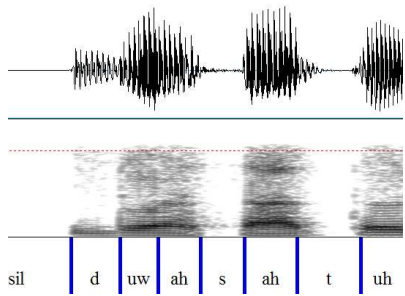

Fig. 7. *subword marking*

*c) Noise data*: in this phase *noise* datas are included into training database. Noise present is a must in online testing. Example of *noise* is shown in SFS display as shown in Figure 8.

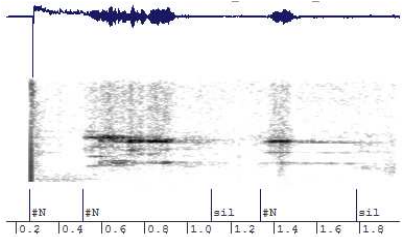

Fig. 8. *Noise marking*

### D. Feature Extraction

Feature extraction to convert waveform to MFCC (Mel Frequency Cepstral Coefficient) format. In this case the raw waveform sound is recorded from the *replay buffer* of HVite or using HSLab program recording. Feature extraction is done by an HTK program called **HCopy**, which is to convert raw PCM sound files with extension **.htk** to feature files with extension **.mfc**. The specific MFCC format used is with the HTK code MFCC_0_D_A. This format is the MFCC format with spectral coefficients to zero, (zeroth coefficient), delta coefficient and acceleration coefficient [8]. Previously pre-processing was done, namely emphasizing with coefficients of 0.97 with windowing of 250 ms (400 samples) consisting of framing, striding of 100ms (160 samples) and overlapping of 150 ms (240 samples). After obtaining MFCC_0_D_A, then proceed with PCA [11], where based on the results of PCA, the number of the required coefficients can be reduced.

### E. Acoustic model creation

In HTK, acoustic models are made using **HInit** and **HRest** and **HERest** programs. HInit initializes parameters using the Viterbi Extraction algorithm. HRest estimates HMM-GMM parameters using the Baum-Welch algorithm for small vocabulary and fixed variance version [1]. HRest is outperformed HERest here because of better performance in noisy environment and complete dataset. This is because HERest is mainly intended to train acoustic speech where the exactly sub-word time boundary transcription is not completely present[1] but only complete sentence patterns that are available. So for complete labelled data, only HInit and HRest is used as shown in figure 9.

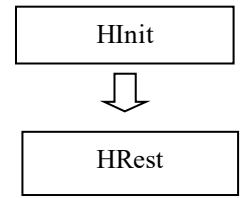

Fig. 9. *HTK programs for creating HMM acoustic model*

### F. Testing

Tests were carried out with the **HVite** program. Hvite can perform testing offline using a recorded database or online directly from the microphone. In this case Bluetooth microphone is used.

### G. Refining

Repair is done by looking at the results of online testing. If in online testing an error occurs in the sentence that is spoken, then the sentence is taken and labelled and added to the database, then retest the sentence again. After that, it is examined whether the previous sentence can be recognized correctly now. The results showed that this method succeeded in correcting the errors of online testing.

Sometimes even though errors data have been entered in the database but still cannot eliminate errors, repairs can be done by looking at the phoneme sequence that is wrongly detected then adding the entry to the dictionary as a word explicitly. At Fig. 10 we show a multi-version entry for spelling NUM_3, whenever there is an entry that is not explicitly added mainly will cause an error.

```
NUM_3        t+ih ih-g+ah g-ah+ah
NUM_3        ah-t+ih t-ih+g ih-g+ah uw-ah
```

Fig. 10. *Multiple word pronunciation accomodation in dictionary*

### H. Deployment

Planting is done on computers with Windows operating systems within the same noise environments between testing and training stages. The software that is planted is **HVite** from HTK. The speaker will pronounce the digits repeatedly and randomly via the Bluetooth microphone input. **HVite** will record and recognize the speech based on the trained HMM-GMM acoustic model. The records kept can be used to refine the system whenever the error occurs.

When applying to real life application like controlling electrical switch for home-based electrical appliance, a threshold is applied to revalidate recognition result. This mechanism is provided by HTK with a parameter *of frame probability*. An upper layer application will execute a command if only the frame probability is higher than a threshold. If the value is below the threshold, the command is discarded.

An example of deploying SVASR application is presented below. This configuration will enable the toggle of electrical switches using voice commands. The main components are:

*1) HTK Speech Recognition server, this is a personal computer or smartphone in which a modified HTK is installed and bluetooth microphone is connected.*

*2) Arduino Uno-Wifi R3 Robodyn, contain Espressif ESP 8266 connected with Atmel Atmega 328p through serial pins. Atmega328p*[14] *is used to control electrical switch while the ESP 8266*[15] *is used to communicate with speech recognition server. The appearance of this board as in Figure-11.*

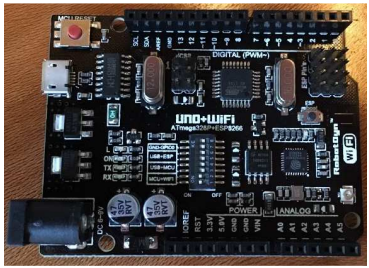

Fig. 11. *Embedded board with Atmega 328p and ESP 8266*

The implementation diagram of the whole system is presented in figure 12:

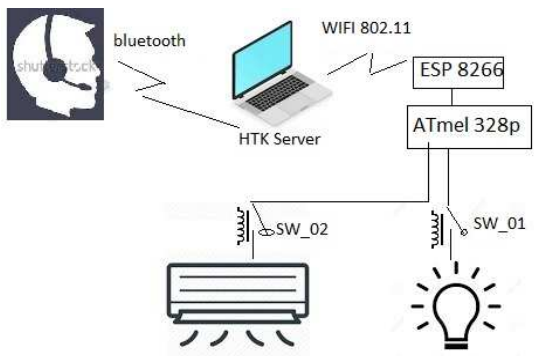

Fig. 12. *Voice based electrical switcher implementation diagram*

## V. DISCUSSION

This procedure using HTK, has a dependency by requiring the same environmental conditions for testing and training. It also requires similarity of speakers and of microphone input devices. The desired development should support more words. It is also expected to minimize or eliminate the need of pre-recording procedures by utilizing the existing global database.

## VI. CONCLUSION

HTK is able to build noise robust automatic speech recognition system in a moderated noisy level environment, especially for small vocabulary system. That is HTK is practically a solution to develop SVASR fast and accurate. When apply for voice-based electrical switch application for home automation in Indonesian language, HTK achieved 100% command recognition rate. In this case the system designed can be uttered to toggle all switches correctly, either by speaking the switch number and also by using the switch meaningful name.

REFERENCES

[1] S. Young, E. Gunnar, G. Mark, T. Hain, and D. Kershaw, "The HTK Book version 3.5 alpha," Cambridge University, 2015.

[2] C. Lopes and F. Perdigão, "Phone Recognition on the TIMIT Database," 2009.

[3] D. P. Lestari, K. Iwano, and S. Furui, "A large vocabulary continuous speech recognition system for Indonesian language," *15th Indones. Sci. Conf. Japan Proc.*, pp. 17–22, 2006.

[4] H. Kokubo, A. Amano, and N. Hataoka, "Robust Speech Recognition for Car Environment Noise," *Electron. Commun. Japan, Part 3*, vol. 85, no. 11, pp. 2190–2197, 2002.

[5] C. D. Soderberg and K. S. Olson, "Illustration of the IPA: Indonesian," *J. Int. Phon. Assoc.*, vol. 38, no. 2, pp. 209–213, 2008.

[6] K. Lee and H.-W. Hon, "Speaker-Independent Phone Recognition Using Hidden Markov Models," *IEEE Trans. Acoust.*, vol. 37, no. 11, pp. 1641–1648, 1989.

[7] M. A. Huckvale, D. M. Brookes, L. T. Dworkin, M. E. Johnson, D. J. Pearce, and L. Whitaker, "The SPAR Speech Filing System," *Eur. Conf. Speech Technol.*, pp. 305–308, 1987.

[8] P. Boersma and V. van Heuven, "Speak and unSpeak with Praat," *Glot Int.*, vol. 5, no. 9–10, pp. 341–347, 2001.

[9] K. John and A. W. Black, "The CMU ARCTIC Speech Databases," in *5th ICSA Speech Synthesis Workshop - Pittsburg*, 2004, pp. 223–224.

[10] S. B. Davis and P. Mermelstein, "Comparison of Parametric Representations for Monosyllabic Word Recognition in Continuously Spoken Sentences," *IEEE Trans. Acoust.*, vol. 28, no. 4, pp. 357–366, 1980.

[11] L. I. Smith, "A tutorial on Principal Components Analysis," 2002.

[12] T. Takiguchi and Y. Ariki, "PCA-Based Speech Enhancement for Distorted Speech Recognition," *J. Multimed.*, vol. 2, no. 5, pp. 13–18, 2007.

[13] L. R. Rabiner, "A Tutorial on Hidden Markov Models and Selected Applications in Speech Recognition," in *Proceedings of*

*the IEEE*, 1989, vol. 77, no. 2, pp. 257–286.

[14] Microchip, "ATmega328P 8-bit AVR Microcontroller with 32K Bytes In-System Programmable Flash DATASHEET."

[15] Espressif Systems, "ESP8266EX Version 6.0 Datasheets," 2018.

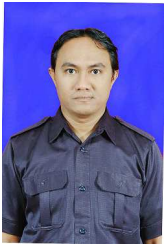

Zulkarnaen Hatala was born in Ambon, on 19 Agustus 1977. He received Sarjana Teknik on Informatics at 2002 and Master Teknik on Telecommunications at 2005. Both degress are from Telkom University, Bandung Indonesia. From 2005 to 2009 he thought computer science and informatics in private universities. Since 2009 he works at Politeknik Negeri Ambon, Indonesia as a lecturer on Electrical Engineering Department.